\documentclass{emulateapj}

\newcommand{\myemail}{quanz@astro.phys.ethz.ch}

\slugcomment{To appear in ApJ Letters}

\shorttitle{NACO/APP 4--micron observations of $\beta$ Pictoris b}
\shortauthors{Quanz et al.}

\begin{document}


\title{First results from VLT NACO Apodizing Phase Plate: \\4--micron images of the exoplanet $\beta$ Pictoris b}


\author{Sascha P. Quanz, Michael R. Meyer}
\affil{Institute for Astronomy, ETH Zurich, Wolfgang-Pauli-Strasse 27, 8093 Zurich, Switzerland}    
\email{\myemail}
\author{Matthew Kenworthy}
\affil{Sterrewacht Leiden, P.O. Box 9513, Niels Bohrweg 2, 2300 RA Leiden, The Netherlands}  
\author{Julien H. V. Girard}
\affil{European Southern Observatory, Alonso de C\'ordova 3107, Vitacura, Cassilla 19001, Santiago, Chile}
\author{Markus Kasper}
\affil{European Southern Observatory, Karl Schwarzschild Strasse, 2, 85748 Garching bei M\"unchen, Germany.}        
\author{Anne-Marie Lagrange}
\affil{Laboratoire dÕAstrophysique, Observatoire de Grenoble, Universit\'e Joseph Fourier, CNRS, BP 53, F-38041 Grenoble, France}    
\author{Daniel Apai}
\affil{Space Telesope Science Institute, 3700 San Martin Drive, Baltimore, MD 21218, USA}
\author{Anthony Boccaletti}
\affil{LESIA, UMR 8109 CNRS, Observatoire de Paris, UPMC, Universit\'e Paris-Diderot, 5 place J. Janssen, 92195 Meudon, France}
\author{Micka\"el Bonnefoy, Gael Chauvin}
\affil{Laboratoire dÕAstrophysique, Observatoire de Grenoble, Universit\'e Joseph Fourier, CNRS, BP 53, F-38041 Grenoble, France}    
\author{Philip M. Hinz}
\affil{Steward Observatory, The University of Arizona, 933 N. Cherry Ave., Tucson, AZ 85721, USA}
\author{Rainer Lenzen}
\affil{Max Planck Institute for Astronomy, K\"onigstuhl 17, 69117 Heidelberg, Germany} 

\altaffiltext{1}{Based on observations collected at the European Organisation for Astronomical Research in the Southern Hemisphere, Chile, under program number 060.A-9800(J).}


\begin{abstract}
Direct imaging of exoplanets requires both high contrast and high spatial resolution. Here, we present the first scientific results obtained with the newly commissioned Apodizing Phase Plate coronagraph (APP) on VLT/NACO. We detected the exoplanet $\beta$ Pictoris b in the narrow band filter centered at 4.05 $\mu$m (NB4.05). The position angle ($209.13^{\circ}\pm2.12^{\circ}$) and the projected separation to its host star ($0.''354\pm0''.012$, i.e., 6.8 $\pm$ 0.2 AU at a distance of 19.3 pc) are in good agreement with the recently presented data from \citet{lagrange2010}. Comparing the observed NB4.05 magnitude of 11.20 $\pm$ 0.23 mag to theoretical atmospheric models we find a best fit with a 7--10 M$_{\rm Jupiter}$ object for an age of 12 Myr, again in agreement with previous estimates. 
Combining our results with published $L'$ photometry we can compare the planet's $[L'-{\rm NB4.05}]$ color to that of cool field dwarfs of higher surface gravity suggesting an effective temperature of $\sim$1700 K. The best fit theoretical model predicts an effective temperature of $\sim$1470 K, but this difference is not significant given our photometric uncertainties. Our results demonstrate the potential of NACO/APP for future planet searches and provides independent confirmation as well as complementary data for $\beta$ Pic b. 
\end{abstract}



\keywords{stars: pre-main sequence --- stars: formation --- planets and satellites: formation --- planets and satellites: detection}


\section{Introduction}


The first images of exoplanets around stars were published in the last two years (HR8799 b, c, d -- \citet{marois2008}; Fomalhaut b -- \citet{kalas2008}; 1RXS J1609-2105 -- \citet{lafreniere2008,lafreniere2010}) following the discovery of a planetary mass companion to the brown dwarf 2MASS1207-3932 in 2005 \citep{chauvin2005}. For the planets around HR8799 follow-up observations at 3--5$\mu$m revealed evidence for non-equilibrium chemistry in the planetary atmospheres \citep{janson2010,hinz2010} 
highlighting the potential of multi-wavelength imagery of extrasolar planetary systems to constrain physical properties in comparison to theoretical atmospheric models. 


\citet{lagrange2009a} detected a planetary mass candidate around the young A-type star $\beta$ Pictoris \citep[A5V, 12$^{+8}_{-4}$ Myr, 19.3$\pm$0.2 pc, 1.75 M$_{\sun}$;][]{zuckerman2001,crifo1997} in $L'$ images from 2003. After a non-detection based on data obtained in early 2009 \citep{lagrange2009b}, \citet{lagrange2010} recently re-detected the planet $\beta$ Pictoris b which had moved on the other side of the star. The authors constrained the semi-major axis to 8--15 AU (the smallest of all imaged exoplanets today) and derived a mass estimate of 9$\pm$3 M$_{\rm Jupiter}$. 

Here, using the newly commissioned apodizing phase plate (APP) on VLT/NACO \citep{kenworthy2010, codona2006}, we confirm the detection of $\beta$ Pictoris b with an independent data set and provide complementary photometry. 

\section{Observations and data reduction}
\begin{figure}[!h]
\centering
\epsscale{1.}
\plotone{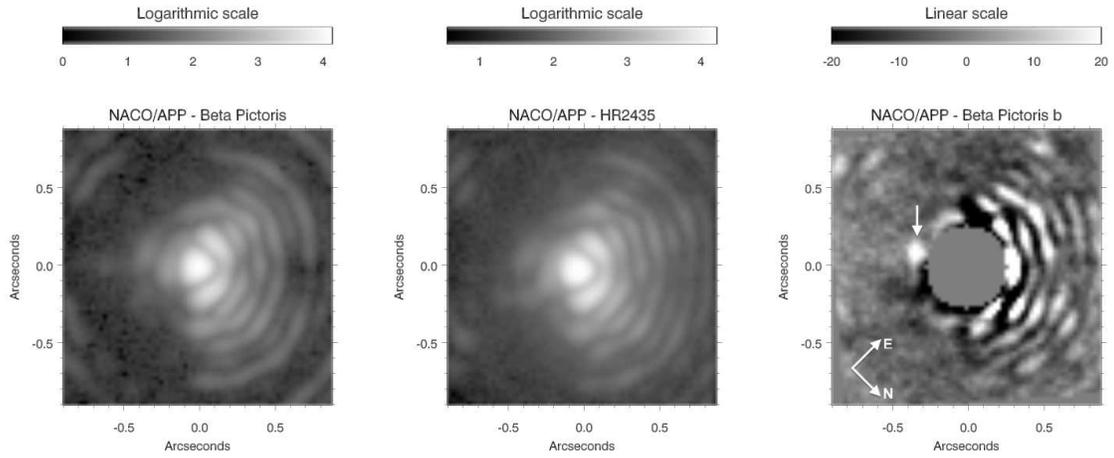}
\caption{\emph{Left:} Median combination of one cube of unsaturated exposures of $\beta$ Pictoris used to determine the photometry. The effect of the APP can be seen in the left side of the PSF where the diffraction rings are effectively suppressed increasing the contrast between 2 and 7 $\lambda/D$. \emph{Middle:} Median combination of one cube of saturated exposures of the PSF reference star HR2435.  \emph{Right:} Median combination of all PSF subtracted science exposures. $\beta$ Pictoris b is indicated by the arrow. The right side suffers from subtraction residuals as does the central region of the PSF which has been masked out. 
\label{finalimage}}
\end{figure}

The data were obtained on 2010-04-03, during the commissioning of the APP with the AO high-resolution camera NACO 
\citep{lenzen2003,rousset2003} mounted on ESO's UT4 on Paranal. The APP is designed to work in the 3--5 $\mu$m range where it enhances the contrast capabilities between $\sim$2--7 $\lambda/D$ on one side of the PSF. We used the L27 camera ($\sim$ 27 mas pixel$^{-1}$) with the visible wavefront sensor. All images were taken in pupil stabilized mode with the NB4.05 filter ($\lambda_{c}=4.05\,\mu$m, $\Delta\lambda=0.02\,\mu$m)\footnote{The APP has been optimized to work with the NB4.05 and IB4.05 filter but it can also be used with the $L'$ filter \citep{kenworthy2010}.}. We used the "cube mode" where all data frames are saved individually. To ensure that no frames were lost we only read out a 512$\times$512 pixel sub-array of the detector. As we knew the position angle where the planet was to be expected \citep{lagrange2009a,lagrange2009b} we rotated the camera by $140^{\circ}$ so that the planet would appear in the "clean", high-contrast side of the APP PSF (Fig.~\ref{finalimage}, left panel).

We obtained six data cubes of $\beta$ Pictoris and directly thereafter six data cubes of the PSF reference star HR2435. HR2435 has already been used in previous studies as reference star for  $\beta$ Pictoris \citep[e.g.,][]{lagrange2009a} as one can obtain data at matching parallactic angles. Also, the star is close in the sky and has a similar brightness providing comparable AO correction. For both sources, each data cube was taken at a slightly different dither position following a  3--point dither pattern which was repeated twice.  
The on-source integration time was 20 minutes each. Table~\ref{observations} summarizes the observations and the observing conditions. We chose to saturate the core of the stellar PSFs but we note that the APP reduces the peak flux in the PSF core by roughly 40\% so that the comparatively long exposure time of 1 sec did not lead to saturation effects outside of the inner $\sim$5 pixels.

For the photometric calibration we also obtained unsaturated images of $\beta$ Pictoris ($\sim$50\% Full Well) prior to the science observations described above. We used the same observing strategy but decreased the detector integration time to 0.2 sec.

The data reduction was done using self-developed IDL routines. The following steps were applied to all three datasets (i.e., $\beta$ Pictoris unsaturated and saturated images, HR2435 saturated images). First, in order to eliminate the sky background emission, we subtracted from each individual frame the corresponding frame from another cube taken at a different dither position (e.g., Cube 2 frame 10 - Cube 3 frame 10) and vice versa. As the first two frames in each cube always showed a higher detector noise level they were disregarded. We also disregarded frames where the AO correction was poor (flux measured in the PSF core less than 50\% compared to the previous frame). Then, bad pixels and cosmic ray hits deviating by more than 3-$\sigma$ in a 5$\times$5 pixels box were replaced with the mean of the surrounding pixels. After this we continued as follows: 
The unsaturated $\beta$ Pictoris images were aligned and cube-wise median combined, yielding six final images on which we performed photometry (Fig.~\ref{finalimage}, left panel). The alignment of the images was done using cross-correlation where the optimal shift between two images was determined with an accuracy of ~0.1 pixel. The same procedure was applied to the saturated HR2435 images yielding six individual reference PSF images (i.e., one per cube; Fig.~\ref{finalimage}, middle panel). The individual frames were not de-rotated to the same field orientation before the combination, so that all static telescope aberrations remained constant throughout the cubes and in the final images.
For the saturated frames of $\beta$ Pictoris we determined the parallactic angle for each individual frame by linear interpolation between the first frame and the final frame in each cube\footnote{Per default, only the parallactic angle at the beginning and at the end of each cube are saved in the fits header in NACO's cube mode.}.
Using cross-correlation, all frames from all cubes were then aligned to the same reference image for which we used the final, median-combined image of the first cube of the unsaturated $\beta$ Pictoris exposures. Thereafter we aligned, scaled and subtracted one of the six final HR2435 reference PSFs from each individual saturated $\beta$ Pictoris frame. The choice of the reference PSF was a trade-off between matching parallactic angle and observing conditions in the individual cubes. The best results (i.e., strongest signal of the planet, least residuals) were obtained using reference PSF 2 for the frames in cubes one and two, PSF 3 for cubes three and four, and PSF 5 for cubes five and six. The scaling was done in the bright, righthand side of the PSF where the diffraction rings were clearly visible. We scaled the reference PSF by minimizing the mean in the subtracted images in a semi-annulus centered on the star with an inner radius of 15 pixels, an outer radius of 27 pixels (i.e., covering the 3$^{\rm rd}$ and 4$^{\rm th}$ Airy ring) and an opening angle of 150$^\circ$. Finally, we derotated all PSF-subtracted frames to match the parallactic angle of the first frame and created a median combined final image.


\section{Results}
\subsection{The detection of $\beta$ Pic b}
The right-hand panel in Fig.~\ref{finalimage} shows the final image of the data reduction process. The exoplanet is clearly detected in the left side of the image. We conducted several tests to exclude the possibility of a false detection: 
(1) We created final images for each individual cube by derotating and combining the respective PSF subtracted images. The planet was detected in the final images of all six cubes. (2) Comparing the position angle of the exoplanet in the final images of the individual cubes revealed the expected rotation introduced by the pupil tracking mode. (3) Speckles can appear along the spider arms holding the secondary mirror of the telescope, but the nearest arm of the telescope spider was $\ga 20^{\circ}$ away from the position of the planet. (4) We implemented a second independent data reduction pipeline based on the LOCI algorithm \citep{lafreniere2007b}. As the small field rotation in our data set did not allow us to construct a reference PSF directly from the $\beta$ Pic images \citep[Angular Differential Imaging; e.g.,][]{marois2006} we used the HR2435 data set. For each $\beta$ Pic frame we constructed a reference PSF based on a linear combination of the final HR2435 images so that the residuals in the high-contrast side of the PSF were minimized. With this approach we also recovered the signal of the exoplanet at the same position.  

The projected separation between star and planet is $0.''354\pm0''.012$ (i.e., 6.8 $\pm$ 0.2 AU at a distance of 19.3 pc) and the position angle is $209.13^{\circ}\pm2.12^{\circ}$ (East of North).  The values are mean values derived from the six final PSF subtracted cube images (see, test (2) above) and the errors include the corresponding standard deviation of the mean and systematic uncertainties from the alignment of the images. The position of the star in the unsaturated reference image used for the alignment and the position of the planet in the six final images were determined by fitting a 2-dimensional Gaussian to the respective source. For our final astrometric numbers we had to rely on the plate scale and field orientation provided in the header of the images as we did not observe an astrometric calibrator. However, potential deviations are expected to be small compared to our errors \citep{lagrange2009a}. 
 Comparing the final numbers to the results from \citet{lagrange2010} we find that the position angle is in very good agreement and the separation appears to have increased as  expected from the planet's orbital motion. However, the short time baseline between our data and the data from \citet{lagrange2010} does not allow us to put any new constraints on the planet's actual orbit. 

\subsection{Photometry and color of $\beta$ Pic b}
To derive the relative photometry between the planet and its star we used an aperture with a radius of 2 pixels and computed the mean flux of $\beta$ Pictoris in the six final unsaturated images and the mean flux of the planet in the six final PSF subtracted cube images. The flux derived from the unsaturated images was scaled to account for the difference in the integration time. Each flux measurement was corrected for residual background emission by measuring and subtracting the mean flux per pixel in a semi-annulus centered on the planet as well as on the star. The semi-annulus covered only the high-contrast side of the objects and had an inner and outer radius of 5 and 8 pixel for the planet and of 18 and 25 pixel for  the star. The contrast between the star and the planet in the NB4.05 filter amounts to $\Delta m_{\rm{4.05}}= 7.75\pm0.23$ mag. The error was derived from error propagation taking into account the standard deviations of the mean of both the flux of the planet and the flux of the unsaturated $\beta$ Pictoris images. 

In the $L'$ filter the contrast between the planet and the star  is $\Delta m_{\rm{L'}}= 7.7\pm0.3$ mag\footnote{This figure was obtained from the October 2009 dataset in \citet{lagrange2010} as well as from the November 2003 dataset in \citet{lagrange2009a}.} which, combined with the stars apparent magnitude of $L'=3.454\pm0.003$ mag \citep{bouchet1991}, translates into an apparent magnitude for the exoplanet of $m_{\rm{L'}}=11.15\pm0.3$ mag. Since an A5V star has an $[L'-M]$ color of $\sim$0.01 mag we can assume that the intrinsic $[L'-{\rm NB4.05}]$ color of $\beta$ Pic is negligible. Based on the observed contrast in the NB4.05 filter the exoplanet's apparent magnitude is then $m_{\rm{NB4.05}}=11.20\pm0.23$ and its color $[L'-{\rm NB4.05}] = -0.05\pm0.38$ mag.

\subsection{Color comparison to cool field dwarfs}
\citet{cushing2008} published NIR spectra for cool L and T field dwarfs that cover the NB4.05 filter but terminate before the long wavelengths cut-off of ESO's $L'$ filter. However, for some of these objects \citet{leggett2002} published $L'$ photometry. To obtain the magnitudes in the NB4.05 filter we used the filter transmission curve and convolved it with the spectra from \citet{cushing2008}. The zero point of the filter was derived from a Kurucz model of an A0V star. Using the $L'$ filter transmission curve we first scaled the Kurucz model to the observed flux density of Vega in the $L'$ filter. For this we assumed $m_{\rm L'}^{{\rm Vega}}=m_{\rm V}^{{\rm Vega}}=0.03$ mag and $F_\nu^{{\rm Vega}}=246.105$ Jy or $F_\lambda^{{\rm Vega}}=5.219\cdot10^{-11}$ W m$^{-2}$ ${\mu}$m$^{-1}$ in the ESO $L'$ filter. Then we measured the flux density of the A0V star in the NB4.05 filter and derived the zero point assuming the star has the same magnitude here. The derived flux densities of the field objects could then be converted into magnitudes and we determined the $[L'-{\rm NB4.05}]$ color for the objects. The error in the magnitude in the NB4.05 filter was derived from computing a minimum and a maximum magnitude by either subtracting or adding the mean error of the spectra in that wavelength range. Whatever resulted in a bigger deviation from the mean magnitude was used as (conservative) error bar. Finally, as the $L'$ photometry from \citet{leggett2002} was done in the MKO photometric system and not in the ESO system we applied a first order correction by subtracting the zero point offset of 0.02 mag\footnote{see, http://www.gemini.edu/sciops/instruments/midir-resources/imaging-calibrations/fluxmagnitude-conversion} from the quoted photometric measurements.

In Fig.~\ref{field_objects} we plot the spectral type of the field objects against their color (black points). Data were available for spectral types L1, L3.5, L7.5, T2 and T5\footnote{Optical spectral types differ slightly from NIR spectral types for 3 sources. However, the outcome of this analysis remains unchanged if we use the NIR spectral types instead.}. Fitting a straight line to the data points we find that the color of $\beta$ Pic b is most similar to that of a field object with a spectral type of L4. However, spectral types from L0 to L7 are consistent with the observed color given the large error bars. 
For field objects around spectral type L4, \citet{cushing2008} found effective temperatures of ~1700 K which is the same temperature one obtains using the spectral type -- temperature relation derived by \citet{golimowski2004}. We note, however, that the surface gravity of the young exoplanet is expected to be much lower than that of the field objects \citep[log $g\approx4.0$ rather than 4.5--5.5 for field L/T dwarfs;][]{cushing2008}.

\subsection{Comparison to atmospheric/evolutionary  models}
\begin{figure}[!b]
\centering
\epsscale{1.}
\plotone{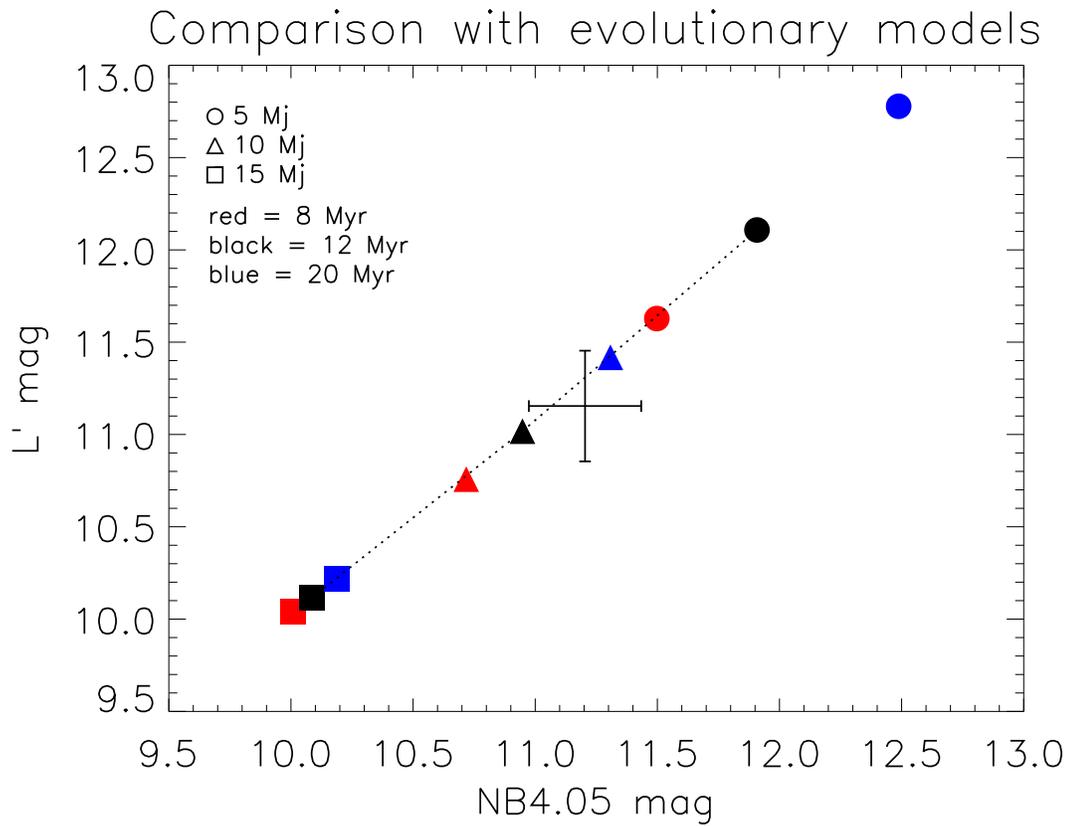}
\caption{The apparent L' and NB4.05 magnitudes of $\beta$ Pic b (black cross) compared to theoretical isochrones based on the DUSTY models from \citet{chabrier2000}. Circles, triangles and boxes denote objects with 5, 10, and 15 M$_{\rm Jupiter}$, respectively. The age of the objects is color-coded with red symbols being 8 Myr, black being 12 Myr, and blue being 20 Myr. The dashed black line shows the 12 Myr isochrone. 
\label{model}}
\end{figure}

In Fig.~\ref{model} we compare the observed magnitudes to theoretical isochrones for low-mass objects based on the DUSTY models from \citet{chabrier2000}. These evolutionary models are based on a spherical collapse model where the objects start initially with very large radii and their internal energy is dominated by contraction. Using a distance of 19.3 pc and assuming an age of 12 Myr we find a mass of 7--10 M$_{\rm Jupiter}$ which agrees with the 6--12 M$_{\rm Jupiter}$ derived from the $L'$ filter. Here, the mass range is only determined by the photometric uncertainties. The most likely mass based on the NB4.05 magnitude is 8 M$_{\rm Jupiter}$ which corresponds to $T_{\rm eff}\approx1470$ K and log $g\approx4.0$. Assuming an age of 20 Myr this mass would increase to 11 M$_{\rm Jupiter}$ with $T_{\rm eff}\approx1425$ K and log $g\approx4.1$, while for 8 Myr we would estimate 6 M$_{\rm Jupiter}$ with $T_{\rm eff}\approx1380$ K and log $g\approx3.9$.

\subsection{Contrast curve analysis}
\begin{figure}[!t]
\centering
\epsscale{1.}
\plotone{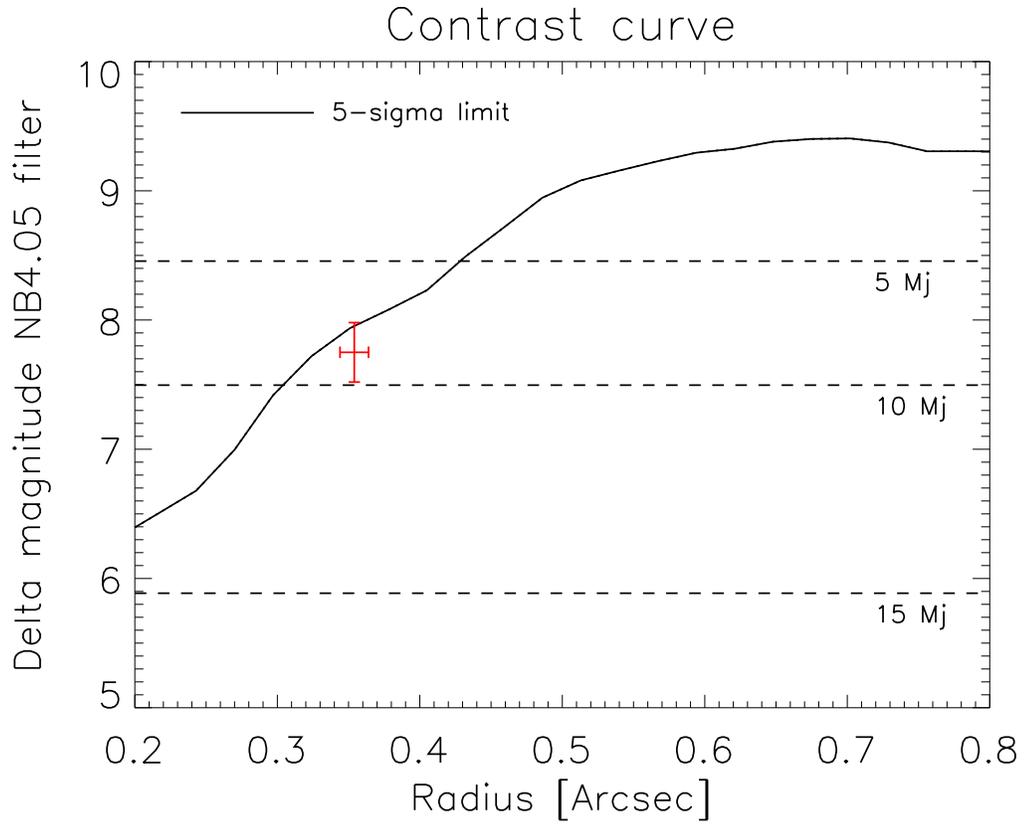}
\caption{5--$\sigma$ Contrast curve derived from the high-contrast side in final PSF-subtracted image shown in the right panel of Fig.~\ref{finalimage}. The dashed horizontal lines correspond to masses as predicted by the DUSTY models \citep{chabrier2000} for an assumed age of 12 Myr. The red cross shows the position of $\beta$ Pic b.
\label{contrast_curve}}
\end{figure}

Using the final image (right panel, Fig.~\ref{finalimage}) we computed a contrast curve showing the detection limit for potential additional companions as a function of radius (Fig.~\ref{contrast_curve}). To derive the curve we computed the standard deviation $\sigma$ of the pixel values in semi-annuli (5 pixel width) as a function of radial distance from the center in the high-contrast side of the image. The signal and the noise of a hypothetical companion can be written as
\begin{equation}
S=n\cdot\eta\cdot\sigma\quad 
\end{equation}
and
\begin{equation}
N\approx\sqrt{n}\cdot\sigma
\end{equation}
with $n$ being the number of pixels in the chosen photometric aperture (here: $n=\pi\cdot2^2\approx12.56$) and $\eta$ being a factor that depends of the chosen S/N ratio one wants to achieve. For a S/N of 5, $\eta$ can be derived from 
\begin{equation}
\frac{S}{N}=5\approx\sqrt{n}\cdot\eta
\end{equation}
The ratio between the signal of a planet with a S/N of 5 (equation (1)) and the mean flux of beta Pictoris computed in a 2 pixel aperture relates to a difference in magnitude which we compute as a function of $\sigma(r)$ (Fig.~\ref{contrast_curve}). As the noise of the residuals is not perfectly Gaussian in all annuli \citep[see also, e.g.,][]{kasper2007} we chose to plot the conservative 5-$\sigma$ limit only. A more sophisticated analysis, i.e., the insertion of fake planets, may result in a somewhat different contrast curve. 

\section{Discussion}
\begin{figure}[!t]
\centering
\epsscale{1.}
\plotone{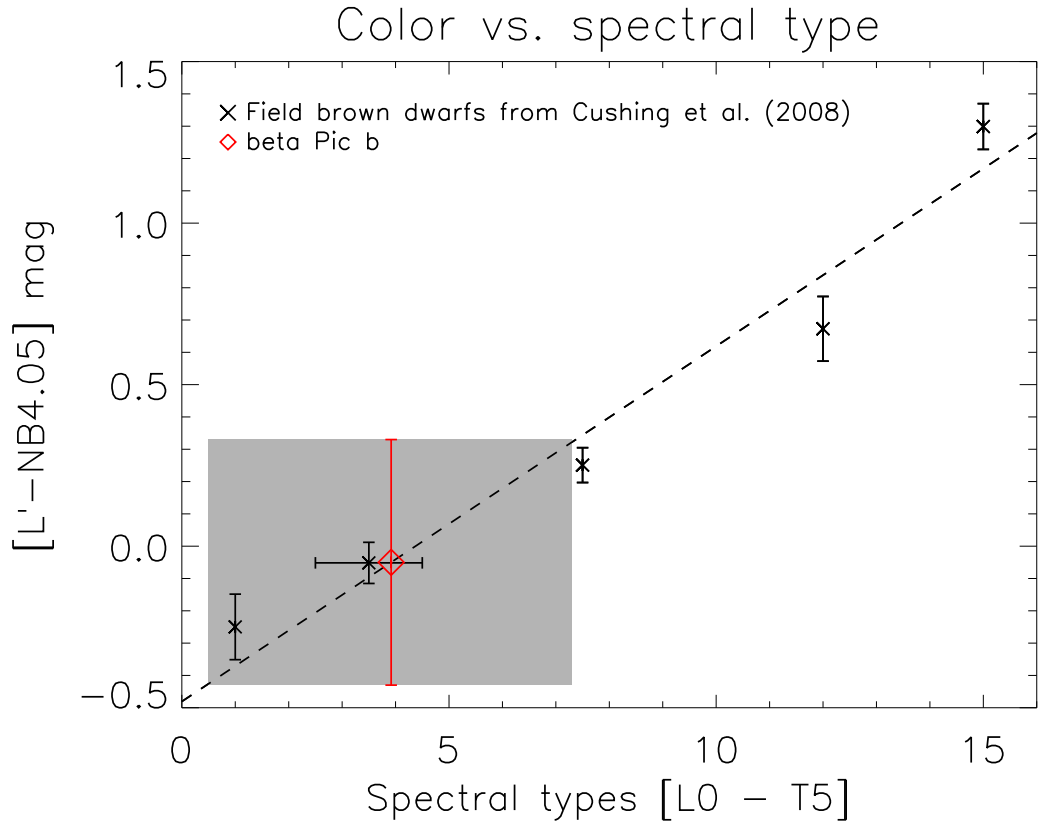}
\caption{The [L' - NB4.05] color as function of spectral type derived from cool field dwarf (black points). The dashed line is a linear fit to the field dwarf data. The position of $\beta$ Pic b on the fitted correlation is shown in red. The error bars are the root-sum-squares of the photometric errors in the individual filters.  The shaded area covers the range of spectral types consistent with the color of the exoplanet given the error bars. 
\label{field_objects}}
\end{figure}

$\beta$ Pictoris b is an exoplanet whose estimated mass, age and separation are in agreement with predictions from core accretion planetary formation models  \citep[see,][and references therein]{lagrange2010}. It is interesting to note that dynamical studies analyzing the sub-structure of the remnant debris disk around $\beta$ Pic predicted the mass and the orbit of the planetary companion before it was imaged \citep[e.g.,][]{mouillet1997,augereau2001}. It was shown that an object with a mass significantly higher than $\sim$10 M$_{\rm Jupiter}$ would create much larger disk asymmetries. Thus, the dynamical analysis provides further support for the mass estimates derived from the observed photometry and the evolutionary models. This in turn led \citet{lagrange2010} to the conclusion that the "cold start" evolutionary models of young, giant planets \citep{fortney2008} are not in agreement with $\beta$ Pic b. These models predict the planet to be significantly less luminous for its age and dynamically predicted mass. Our new flux point at 4.05 $\mu$m confirms the brightness of the source and supports the argument from \citet{lagrange2010}. However, given the comparatively large uncertainties in the observed fluxes we refrain from undertaking a more comprehensive comparison to additional sets of evolutionary and atmospheric models \citep[e.g.,][]{burrows2006}. For instance for an assumed age of 12 Myr the models from \citet{chabrier2000} predict [$L'-{\rm NB4.05}$] $\le$ 0.3 mag for $\emph{all}$ objects between 4 and 100 M$_{\rm Jupiter}$. Also, the apparent difference in effective temperature of $\beta$ Pic b derived in sections 3.3. and 3.4. is not statistically significant. Given the error bars in Fig.~\ref{field_objects}, $T_{\rm eff}$ as high as $\sim$1700 K is still consistent  with the observed magnitudes within 2--$\sigma$. 

Future observations at different wavelengths will help us to better constrain the atmospheric parameters and composition of $\beta$ Pic b and to check whether the preliminary result, that the models predict a lower $T_{\rm eff}$ than the comparison to field dwarfs, persists. A frequent monitoring of the exoplanet's position will eventually allow us to determine the orbital parameters with higher accuracy.



\acknowledgments
This research has made use of the SIMBAD database, operated at CDS, Strasbourg, France.  We are indebted to Michael Cushing, Isabelle Baraffe, Udo Wehmeier and the ESO staff on Paranal, in particular Jared O'Neal, for their support. We also thank the referee for useful comments and suggestions.



{\it Facilities:} \facility{VLT:Yepun (NACO)}

\clearpage

\begin{deluxetable}{lll}
\tablecaption{Summary of observations on 2010-04-03
\label{observations}}           
\tablewidth{0pt}
\tablehead{
\colhead{Parameter} & \colhead{$\beta$ Pictoris} & \colhead{HR2435} 
}
\startdata
UT start & 00:39:31.1 & 01:34:21.3 \\
NDIT $\times$ DIT\tablenotemark{a}  & 200 $\times$ 1 s & 200 $\times$ 1 s\\
NINT\tablenotemark{b} & 6 & 6 \\
Parallactic angle start & 69.469$^{\circ}$ & 69.451$^{\circ}$ \\
Parallactic angle end & 74.492$^{\circ}$ & 74.499$^{\circ}$ \\
Airmass & 1.38--1.44 & 1.42--1.48 \\
Typical DIMM seeing &  0.70--0.90$''$ & 0.60--0.75$''$ \\
$\langle EC \rangle$\tablenotemark{c} & 29.5--44.4 \% &  22.7--42.9 \%\\
$\langle \tau_0 \rangle$\tablenotemark{d} & 4.3--7.9 ms& 4.0--7.8 ms\\
\enddata
\tablenotetext{a}{NDIT = Number of detector integration times (i.e., number of individual frames); DIT = Detector integration time (i.e., single frame exposure time)}
\tablenotetext{b}{NINT = Number of integrations or data cubes (2 $\times$ 3 dither positions = 6 data cubes)}
\tablenotetext{c}{Average value of the coherent energy of the PSF per data cube. Calculated by the Real Time Computer of the AO system.}
\tablenotetext{d}{Average value of the coherence time of the atmosphere per data cube. Calculated by the Real Time Computer of the AO system.}

\end{deluxetable}

\clearpage

\clearpage

\clearpage

\clearpage

\begin{thebibliography}{32}
\expandafter\ifx\csname natexlab\endcsname\relax\def\natexlab#1{#1}\fi

\bibitem[{{Augereau} {et~al.}(2001){Augereau}, {Nelson}, {Lagrange},
  {Papaloizou}, \& {Mouillet}}]{augereau2001}
{Augereau}, J.~C., {Nelson}, R.~P., {Lagrange}, A.~M., {Papaloizou}, J.~C.~B.,
  \& {Mouillet}, D. 2001, \aap, 370, 447

\bibitem[{{Bonomo} {et~al.}(2010){Bonomo}, {Santerne}, {Alonso}, {Gazzano},
  {Havel}, {Aigrain}, {Auvergne}, {Baglin}, {Barbieri}, {Barge}, {Benz},
  {Bord{\'e}}, {Bouchy}, {Bruntt}, {Cabrera}, {Cameron}, {Carone}, {Carpano},
  {Csizmadia}, {Deleuil}, {Deeg}, {Dvorak}, {Erikson}, {Ferraz-Mello},
  {Fridlund}, {Gandolfi}, {Gillon}, {Guenther}, {Guillot}, {Hatzes},
  {H{\'e}brard}, {Jorda}, {Lammer}, {Lanza}, {L{\'e}ger}, {Llebaria}, {Mayor},
  {Mazeh}, {Moutou}, {Ollivier}, {P{\"a}tzold}, {Pepe}, {Queloz}, {Rauer},
  {Rouan}, {Samuel}, {Schneider}, {Tingley}, {Udry}, \&
  {Wuchterl}}]{bonomo2010}
{Bonomo}, A.~S., {Santerne}, A., {Alonso}, R., {Gazzano}, J., {Havel}, M.,
  {Aigrain}, S., {Auvergne}, M., {Baglin}, A., {Barbieri}, M., {Barge}, P.,
  {Benz}, W., {Bord{\'e}}, P., {Bouchy}, F., {Bruntt}, H., {Cabrera}, J.,
  {Cameron}, A.~C., {Carone}, L., {Carpano}, S., {Csizmadia}, S., {Deleuil},
  M., {Deeg}, H.~J., {Dvorak}, R., {Erikson}, A., {Ferraz-Mello}, S.,
  {Fridlund}, M., {Gandolfi}, D., {Gillon}, M., {Guenther}, E., {Guillot}, T.,
  {Hatzes}, A., {H{\'e}brard}, G., {Jorda}, L., {Lammer}, H., {Lanza}, A.~F.,
  {L{\'e}ger}, A., {Llebaria}, A., {Mayor}, M., {Mazeh}, T., {Moutou}, C.,
  {Ollivier}, M., {P{\"a}tzold}, M., {Pepe}, F., {Queloz}, D., {Rauer}, H.,
  {Rouan}, D., {Samuel}, B., {Schneider}, J., {Tingley}, B., {Udry}, S., \&
  {Wuchterl}, G. 2010, ArXiv e-prints

\bibitem[{{Bouchet} {et~al.}(1991){Bouchet}, {Schmider}, \&
  {Manfroid}}]{bouchet1991}
{Bouchet}, P., {Schmider}, F.~X., \& {Manfroid}, J. 1991, \aaps, 91, 409

\bibitem[{{Burrows} {et~al.}(2006){Burrows}, {Sudarsky}, \&
  {Hubeny}}]{burrows2006}
{Burrows}, A., {Sudarsky}, D., \& {Hubeny}, I. 2006, \apj, 640, 1063

\bibitem[{{Chabrier} {et~al.}(2000){Chabrier}, {Baraffe}, {Allard}, \&
  {Hauschildt}}]{chabrier2000}
{Chabrier}, G., {Baraffe}, I., {Allard}, F., \& {Hauschildt}, P. 2000, \apj,
  542, 464

\bibitem[{{Chauvin} {et~al.}(2005){Chauvin}, {Lagrange}, {Dumas}, {Zuckerman},
  {Mouillet}, {Song}, {Beuzit}, \& {Lowrance}}]{chauvin2005}
{Chauvin}, G., {Lagrange}, A., {Dumas}, C., {Zuckerman}, B., {Mouillet}, D.,
  {Song}, I., {Beuzit}, J., \& {Lowrance}, P. 2005, \aap, 438, L25

\bibitem[{{Codona} {et~al.}(2006){Codona}, {Kenworthy}, {Hinz}, {Angel}, \&
  {Woolf}}]{codona2006}
{Codona}, J.~L., {Kenworthy}, M.~A., {Hinz}, P.~M., {Angel}, J.~R.~P., \&
  {Woolf}, N.~J. 2006, in Society of Photo-Optical Instrumentation Engineers
  (SPIE) Conference Series, Vol. 6269, Society of Photo-Optical Instrumentation
  Engineers (SPIE) Conference Series

\bibitem[{{Crifo} {et~al.}(1997){Crifo}, {Vidal-Madjar}, {Lallement}, {Ferlet},
  \& {Gerbaldi}}]{crifo1997}
{Crifo}, F., {Vidal-Madjar}, A., {Lallement}, R., {Ferlet}, R., \& {Gerbaldi},
  M. 1997, \aap, 320, L29

\bibitem[{{Cushing} {et~al.}(2008){Cushing}, {Marley}, {Saumon}, {Kelly},
  {Vacca}, {Rayner}, {Freedman}, {Lodders}, \& {Roellig}}]{cushing2008}
{Cushing}, M.~C., {Marley}, M.~S., {Saumon}, D., {Kelly}, B.~C., {Vacca},
  W.~D., {Rayner}, J.~T., {Freedman}, R.~S., {Lodders}, K., \& {Roellig}, T.~L.
  2008, \apj, 678, 1372

\bibitem[{{Dong} {et~al.}(2009){Dong}, {Bond}, {Gould}, {Koz{\l}owski},
  {Miyake}, {Gaudi}, {Bennett}, {Abe}, {Gilmore}, {Fukui}, {Furusawa},
  {Hearnshaw}, {Itow}, {Kamiya}, {Kilmartin}, {Korpela}, {Lin}, {Ling},
  {Masuda}, {Matsubara}, {Muraki}, {Nagaya}, {Ohnishi}, {Okumura}, {Perrott},
  {Rattenbury}, {Saito}, {Sako}, {Sato}, {Skuljan}, {Sullivan}, {Sumi},
  {Sweatman}, {Tristram}, {Yock}, {The MOA Collaboration}, {Bolt}, {Christie},
  {DePoy}, {Han}, {Janczak}, {Lee}, {Mallia}, {McCormick}, {Monard}, {Maury},
  {Natusch}, {Park}, {Pogge}, {Santallo}, {Stanek}, {The {$\mu$}FUN
  Collaboration}, {Udalski}, {Kubiak}, {Szyma{\'n}ski}, {Pietrzy{\'n}ski},
  {Soszy{\'n}ski}, {Szewczyk}, {Wyrzykowski}, {Ulaczyk}, \& {The OGLE
  Collaboration}}]{dong2009}
{Dong}, S., {Bond}, I.~A., {Gould}, A., {Koz{\l}owski}, S., {Miyake}, N.,
  {Gaudi}, B.~S., {Bennett}, D.~P., {Abe}, F., {Gilmore}, A.~C., {Fukui}, A.,
  {Furusawa}, K., {Hearnshaw}, J.~B., {Itow}, Y., {Kamiya}, K., {Kilmartin},
  P.~M., {Korpela}, A., {Lin}, W., {Ling}, C.~H., {Masuda}, K., {Matsubara},
  Y., {Muraki}, Y., {Nagaya}, M., {Ohnishi}, K., {Okumura}, T., {Perrott},
  Y.~C., {Rattenbury}, N., {Saito}, T., {Sako}, T., {Sato}, S., {Skuljan}, L.,
  {Sullivan}, D.~J., {Sumi}, T., {Sweatman}, W., {Tristram}, P.~J., {Yock},
  P.~C.~M., {The MOA Collaboration}, {Bolt}, G., {Christie}, G.~W., {DePoy},
  D.~L., {Han}, C., {Janczak}, J., {Lee}, C., {Mallia}, F., {McCormick}, J.,
  {Monard}, B., {Maury}, A., {Natusch}, T., {Park}, B., {Pogge}, R.~W.,
  {Santallo}, R., {Stanek}, K.~Z., {The {$\mu$}FUN Collaboration}, {Udalski},
  A., {Kubiak}, M., {Szyma{\'n}ski}, M.~K., {Pietrzy{\'n}ski}, G.,
  {Soszy{\'n}ski}, I., {Szewczyk}, O., {Wyrzykowski}, {\L}., {Ulaczyk}, K., \&
  {The OGLE Collaboration}. 2009, \apj, 698, 1826

\bibitem[{{Fortney} {et~al.}(2008){Fortney}, {Marley}, {Saumon}, \&
  {Lodders}}]{fortney2008}
{Fortney}, J.~J., {Marley}, M.~S., {Saumon}, D., \& {Lodders}, K. 2008, \apj,
  683, 1104

\bibitem[{{Golimowski} {et~al.}(2004){Golimowski}, {Leggett}, {Marley}, {Fan},
  {Geballe}, {Knapp}, {Vrba}, {Henden}, {Luginbuhl}, {Guetter}, {Munn},
  {Canzian}, {Zheng}, {Tsvetanov}, {Chiu}, {Glazebrook}, {Hoversten},
  {Schneider}, \& {Brinkmann}}]{golimowski2004}
{Golimowski}, D.~A., {Leggett}, S.~K., {Marley}, M.~S., {Fan}, X., {Geballe},
  T.~R., {Knapp}, G.~R., {Vrba}, F.~J., {Henden}, A.~A., {Luginbuhl}, C.~B.,
  {Guetter}, H.~H., {Munn}, J.~A., {Canzian}, B., {Zheng}, W., {Tsvetanov},
  Z.~I., {Chiu}, K., {Glazebrook}, K., {Hoversten}, E.~A., {Schneider}, D.~P.,
  \& {Brinkmann}, J. 2004, \aj, 127, 3516

\bibitem[{{Hinz} {et~al.}(2010){Hinz}, {Rodigas}, {Kenworthy}, {Sivanandam},
  {Heinze}, {Mamajek}, \& {Meyer}}]{hinz2010}
{Hinz}, P.~M., {Rodigas}, T.~J., {Kenworthy}, M.~A., {Sivanandam}, S.,
  {Heinze}, A.~N., {Mamajek}, E.~E., \& {Meyer}, M.~R. 2010, \apj, 716, 417

\bibitem[{{Janson} {et~al.}(2010){Janson}, {Bergfors}, {Goto}, {Brandner}, \&
  {Lafreni{\`e}re}}]{janson2010}
{Janson}, M., {Bergfors}, C., {Goto}, M., {Brandner}, W., \& {Lafreni{\`e}re},
  D. 2010, \apjl, 710, L35

\bibitem[{{Kalas} {et~al.}(2008){Kalas}, {Graham}, {Chiang}, {Fitzgerald},
  {Clampin}, {Kite}, {Stapelfeldt}, {Marois}, \& {Krist}}]{kalas2008}
{Kalas}, P., {Graham}, J.~R., {Chiang}, E., {Fitzgerald}, M.~P., {Clampin}, M.,
  {Kite}, E.~S., {Stapelfeldt}, K., {Marois}, C., \& {Krist}, J. 2008, Science,
  322, 1345

\bibitem[{{Kasper} {et~al.}(2007){Kasper}, {Apai}, {Janson}, \&
  {Brandner}}]{kasper2007}
{Kasper}, M., {Apai}, D., {Janson}, M., \& {Brandner}, W. 2007, \aap, 472, 321

\bibitem[{{Kenworthy} {et~al.}(2007){Kenworthy}, {Codona}, {Hinz}, {Angel},
  {Heinze}, \& {Sivanandam}}]{kenworthy2007}
{Kenworthy}, M.~A., {Codona}, J.~L., {Hinz}, P.~M., {Angel}, J.~R.~P.,
  {Heinze}, A., \& {Sivanandam}, S. 2007, \apj, 660, 762

\bibitem[{{Kenworthy} {et~al.}(2010){Kenworthy}, {Quanz}, {Meyer}, {Kasper},
  {Lenzen}, {Codona}, {Girard}, \& {Hinz}}]{kenworthy2010}
{Kenworthy}, M.~A., {Quanz}, S.~P., {Meyer}, M.~R., {Kasper}, M.~E., {Lenzen},
  R., {Codona}, J.~L., {Girard}, J.~H.~V., \& {Hinz}, P.~M. 2010, ArXiv
  e-prints

\bibitem[{{Lafreni{\`e}re} {et~al.}(2008){Lafreni{\`e}re}, {Jayawardhana}, \&
  {van Kerkwijk}}]{lafreniere2008}
{Lafreni{\`e}re}, D., {Jayawardhana}, R., \& {van Kerkwijk}, M.~H. 2008, \apjl,
  689, L153

\bibitem[{{Lafreni{\`e}re} {et~al.}(2010){Lafreni{\`e}re}, {Jayawardhana}, \&
  {van Kerkwijk}}]{lafreniere2010}
---. 2010, ArXiv e-prints

\bibitem[{{Lafreni{\`e}re} {et~al.}(2007){Lafreni{\`e}re}, {Marois}, {Doyon},
  {Nadeau}, \& {Artigau}}]{lafreniere2007b}
{Lafreni{\`e}re}, D., {Marois}, C., {Doyon}, R., {Nadeau}, D., \& {Artigau},
  {\'E}. 2007, \apj, 660, 770

\bibitem[{{Lagrange} {et~al.}(2009{\natexlab{a}}){Lagrange}, {Gratadour},
  {Chauvin}, {Fusco}, {Ehrenreich}, {Mouillet}, {Rousset}, {Rouan}, {Allard},
  {Gendron}, {Charton}, {Mugnier}, {Rabou}, {Montri}, \&
  {Lacombe}}]{lagrange2009a}
{Lagrange}, A., {Gratadour}, D., {Chauvin}, G., {Fusco}, T., {Ehrenreich}, D.,
  {Mouillet}, D., {Rousset}, G., {Rouan}, D., {Allard}, F., {Gendron}, {\'E}.,
  {Charton}, J., {Mugnier}, L., {Rabou}, P., {Montri}, J., \& {Lacombe}, F.
  2009{\natexlab{a}}, \aap, 493, L21

\bibitem[{{Lagrange} {et~al.}(2009{\natexlab{b}}){Lagrange}, {Kasper},
  {Boccaletti}, {Chauvin}, {Gratadour}, {Fusco}, {Ehrenreich}, {Apai},
  {Mouillet}, \& {Rouan}}]{lagrange2009b}
{Lagrange}, A., {Kasper}, M., {Boccaletti}, A., {Chauvin}, G., {Gratadour}, D.,
  {Fusco}, T., {Ehrenreich}, D., {Apai}, D., {Mouillet}, D., \& {Rouan}, D.
  2009{\natexlab{b}}, \aap, 506, 927

\bibitem[{Lagrange {et~al.}(2010)Lagrange, Bonnefoy, Chauvin, Apai, Ehrenreich,
  Boccaletti, Gratadour, Rouan, Mouillet, Lacour, \& Kasper}]{lagrange2010}
Lagrange, A.-M., Bonnefoy, M., Chauvin, G., Apai, D., Ehrenreich, D.,
  Boccaletti, A., Gratadour, D., Rouan, D., Mouillet, D., Lacour, S., \&
  Kasper, M. 2010, Science, science.1187187

\bibitem[{{Leggett} {et~al.}(2002){Leggett}, {Golimowski}, {Fan}, {Geballe},
  {Knapp}, {Brinkmann}, {Csabai}, {Gunn}, {Hawley}, {Henry}, {Hindsley},
  {Ivezi{\'c}}, {Lupton}, {Pier}, {Schneider}, {Smith}, {Strauss}, {Uomoto}, \&
  {York}}]{leggett2002}
{Leggett}, S.~K., {Golimowski}, D.~A., {Fan}, X., {Geballe}, T.~R., {Knapp},
  G.~R., {Brinkmann}, J., {Csabai}, I., {Gunn}, J.~E., {Hawley}, S.~L.,
  {Henry}, T.~J., {Hindsley}, R., {Ivezi{\'c}}, {\v Z}., {Lupton}, R.~H.,
  {Pier}, J.~R., {Schneider}, D.~P., {Smith}, J.~A., {Strauss}, M.~A.,
  {Uomoto}, A., \& {York}, D.~G. 2002, \apj, 564, 452

\bibitem[{{Lenzen} {et~al.}(2003){Lenzen}, {Hartung}, {Brandner}, {Finger},
  {Hubin}, {Lacombe}, {Lagrange}, {Lehnert}, {Moorwood}, \&
  {Mouillet}}]{lenzen2003}
{Lenzen}, R., {Hartung}, M., {Brandner}, W., {Finger}, G., {Hubin}, N.~N.,
  {Lacombe}, F., {Lagrange}, A., {Lehnert}, M.~D., {Moorwood}, A.~F.~M., \&
  {Mouillet}, D. 2003, in Society of Photo-Optical Instrumentation Engineers
  (SPIE) Conference Series, Vol. 4841, Society of Photo-Optical Instrumentation
  Engineers (SPIE) Conference Series, ed. {M.~Iye \& A.~F.~M.~Moorwood},
  944--952

\bibitem[{{Marois} {et~al.}(2006){Marois}, {Lafreni{\`e}re}, {Doyon},
  {Macintosh}, \& {Nadeau}}]{marois2006}
{Marois}, C., {Lafreni{\`e}re}, D., {Doyon}, R., {Macintosh}, B., \& {Nadeau},
  D. 2006, \apj, 641, 556

\bibitem[{{Marois} {et~al.}(2008){Marois}, {Macintosh}, {Barman}, {Zuckerman},
  {Song}, {Patience}, {Lafreni{\`e}re}, \& {Doyon}}]{marois2008}
{Marois}, C., {Macintosh}, B., {Barman}, T., {Zuckerman}, B., {Song}, I.,
  {Patience}, J., {Lafreni{\`e}re}, D., \& {Doyon}, R. 2008, Science, 322, 1348

\bibitem[{{Mouillet} {et~al.}(1997){Mouillet}, {Larwood}, {Papaloizou}, \&
  {Lagrange}}]{mouillet1997}
{Mouillet}, D., {Larwood}, J.~D., {Papaloizou}, J.~C.~B., \& {Lagrange}, A.~M.
  1997, \mnras, 292, 896

\bibitem[{{Rousset} {et~al.}(2003){Rousset}, {Lacombe}, {Puget}, {Hubin},
  {Gendron}, {Fusco}, {Arsenault}, {Charton}, {Feautrier}, {Gigan}, {Kern},
  {Lagrange}, {Madec}, {Mouillet}, {Rabaud}, {Rabou}, {Stadler}, \&
  {Zins}}]{rousset2003}
{Rousset}, G., {Lacombe}, F., {Puget}, P., {Hubin}, N.~N., {Gendron}, E.,
  {Fusco}, T., {Arsenault}, R., {Charton}, J., {Feautrier}, P., {Gigan}, P.,
  {Kern}, P.~Y., {Lagrange}, A., {Madec}, P., {Mouillet}, D., {Rabaud}, D.,
  {Rabou}, P., {Stadler}, E., \& {Zins}, G. 2003, in Society of Photo-Optical
  Instrumentation Engineers (SPIE) Conference Series, Vol. 4839, Society of
  Photo-Optical Instrumentation Engineers (SPIE) Conference Series, ed.
  {P.~L.~Wizinowich \& D.~Bonaccini}, 140--149

\bibitem[{{Udry} \& {Santos}(2007)}]{udrysantos2007}
{Udry}, S. \& {Santos}, N.~C. 2007, \araa, 45, 397

\bibitem[{{Zuckerman} {et~al.}(2001){Zuckerman}, {Song}, {Bessell}, \&
  {Webb}}]{zuckerman2001}
{Zuckerman}, B., {Song}, I., {Bessell}, M.~S., \& {Webb}, R.~A. 2001, \apjl,
  562, L87

\end{thebibliography}
\end{document}